\documentclass[twocolumn,showpacs,preprintnumbers,amsmath,amssymb]{revtex4}

\usepackage{graphicx}  
\usepackage{dcolumn}   
\usepackage{bm}        

\begin{document}
\preprint{v 1.55}

\title{Wavepacket reconstruction via local dynamics in
  a parabolic lattice}

\author{Quentin Thommen}
\author{V{\'e}ronique Zehnl{\'e}}
\author{Jean Claude Garreau}
 \affiliation{Laboratoire de Physique des Lasers, Atomes et Mol{\'e}cules\\
  Centre d'Etudes et de Recherches Laser et Applications,
Universit\'{e} des Sciences et Technologies de Lille, \\F-59655
Villeneuve d'Ascq Cedex, France}

\homepage{http://www.phlam.univ-lille1.fr}

\date{\today}

\begin{abstract}
We study the dynamics of a wavepacket
in a potential formed by the sum of a periodic lattice and
of a parabolic potential. The dynamics of the wavepacket 
is essentially a superposition of ``local Bloch oscillations'',
whose frequency is
proportional to the local slope of the parabolic potential. 
We show that the amplitude and the phase of the
Fourier transform of a signal characterizing this dynamics
contains information about the amplitude and the phase
of the wavepacket at a given lattice site.
Hence, {\em complete} reconstruction of the the wavepacket
in the real space can be performed from the
study of the dynamics of the system.
\end{abstract}

\maketitle

The reconstruction of the wavefunction of a quantum system in the real space
from experimentally accessible measurements is seldom possible. Even when
the spatial probability distribution is accessible, phase measurements in
general require the use of a completely different, and often incompatible,
technique. This is essentially due to the fact that a measurement of the
position of a quantum system is usually accompanied by a wavepacket
reduction that erases the phase information.
In a one-atom Magneto-optical trap \cite{ref:OneAtom} (prepared in a
reproductive way), repeated measurements of the fluorescence light allow
reconstruction of the atom probability distribution with a precision of the
order of the wavelength of the detected light, whereas a measurement
of wavepacket phases needs a specific atom-interferometry setup.

In contrast, the quantum dynamics of a system is
highly sensitive to both the amplitude and the phase of the
wavepacket. The idea at the root of the present letter is to use
information on the dynamics of the system to determine simultaneously
both these quantities, with a spatial resolution at least comparable to that
of the methods mentioned in the preceding paragraph. The dynamics
exploited here is related to the so-called ``Bloch oscillation'' (BO)
observed in the motion of a quantum particle in a spatially periodic lattice
to which a constant slope is added, forming a ``tilted lattice''.
In such a potential, instead of climbing down the slope of the
potential, the wavepacket performs  bounded periodic oscillations
in real and momentum spaces. The origin of this
oscillatory motion can be retraced to quantum
interference among pieces of the original wave packet that
are transmitted and reflected by the successive potential barriers of
the lattice, in much analogy with a chain of optical
interferometers with a linearly varying effective index \cite{ref:OptBO}.
The coherence of the wavepacket is thus a pre-requisite to the
observation of the BO. The frequency of oscillation (the ``Bloch
frequency'') corresponds to the energy shift from one
lattice well to the next, and thus to the slope of the potential. This
behavior, theoretically predicted in 1934 \cite{ref:Zener}, has
been observed experimentally in solid-state superlattices \cite
{ref:SolidStateBO}, with cold atoms \cite{ref:BlochOsc}, 
with Bose-Einstein condensates \cite{ref:BECBO}, and also with photons,
in arrays of optical waveguides \cite{ref:OptBO}. A theoretical study
of the BO in an frame close to that adopted here can be found in
Ref.~\cite{ref:WSbasis}.

Let us consider a quantum system, the ``atom'', of mass $M$,
placed in an one-dimensional
potential formed by a spatially-periodic lattice of period $d$,
to which a parabolic term $ax^{2}/2$ is added. With
cold-atoms, this can be obtained by using a far-detuned standing wave which
creates a sinusoidal lattice \cite{ref:BlochOsc,ref:BECBO,ref:ExpWannierStark},
superposed to an independent, focalized laser beam,
producing a parabolic term (a laser-atom detuning much greater than
the natural width of the concerned electronic transition reduces spontaneous
emission, insuring the system to be conservative). The Hamiltonian of
the system is 
\begin{equation}
  H=-\frac{1}{2M^*}
  \frac{\partial^2}{\partial x^2}-V_{0}\cos (2\pi x)+{\frac{x^{2}}{2}}
  \label{eq:H}
\end{equation}
where we introduced dimensionless units in which distances
are measured in units of
the lattice period $d$, energy is measured in units of $ad^{2}$ [or,
equivalently, time is measured in units of $\hbar /(ad^{2}$)],
$M^* \equiv Mad^4/\hbar^2$ is a reduced mass, and $\hbar=1$. Without loss of
generality, we shall use a sinusoidal lattice.

\begin{figure}
\includegraphics{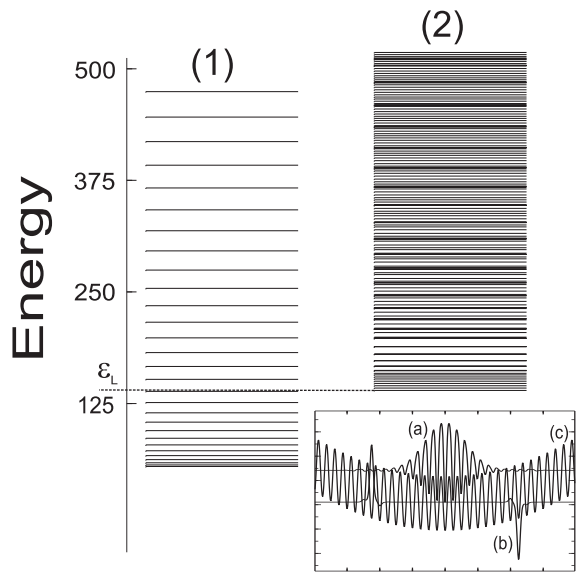} 
\caption{\label{fig:Spectrum} Energy spectrum of the Hamiltonian 
Eq.~(\protect\ref{eq:H}). Column (1) represents the energies of
localized states, with a parabolic distribution. Column (2)
displays the energies of the delocalized states. The inset displays
the wavefunction's modulus for a typical delocalized state (a), and a
(anti-symmetric) localized state (b), compared to the parabolic lattice (c).
$V_0 = 90$.}
\end{figure}

The energy spectrum of the Hamiltonian Eq.~(\ref{eq:H}) is plotted in
Fig.~\ref{fig:Spectrum} and shows two distinct kinds of states.
For energies below a threshold value $\varepsilon _{L}$, the
corresponding eigenstates are
localized in  symmetric lattice wells $-n$ and $n$ (where the lattice
site at $n=0$ is tied to the symmetry axis of the parabola). For each possible
value of the energy, one finds a symmetric and an anti-symmetric
quasi-degenerated states,
which are noted resp. $\phi _{n}^{S}(x)$ and $\phi _{n}^{A}(x)$.
We suppose that the depth of the
lattice is such that there are only one symmetric and one antisymmetric
state for each couple of lattice wells $n$ and $-n$
($V_0 = 90$). The corresponding
eigenenergies are noted $n^{2}/2+\delta _{n}^{S}$ and $n^{2}/2+\delta
_{n}^{A}$. For energies {\em above} $\varepsilon _{L}$, one finds
both localized and delocalized states. At high energies,
the latter are similar to the eigenstates of the
harmonic oscillator and their eigenenergies approximately display a
linear dependence on $n$ (although this is not easily seen in
Fig.~\ref{fig:Spectrum}).

We suppose that the atom wavepacket has been prepared in a
superposition of {\em localized} states only. For high enough lattice
amplitude $V_{0},$ one can neglect the energy shifts $\delta _{n}^{S}$
and $\delta_{n}^{A}$ \cite{note:quasi-degeneracy},
and consider as a new
basis the left and right quasi-eigenstates $\phi _{\pm n}(x) \equiv \left[ \phi
_{n}^{S}(x)\pm \phi _{n}^{A}(x)\right] /\sqrt{2}$, which are localized in the
lattice site $\pm n$ with energy $\epsilon_n = n^2/2$ 
(up to an additive constant). Moreover, the
localized eigenstates $\phi _{n}(x)$ are almost invariant under a
translation by an integer number of steps of the lattice: $\phi
_{n}(x)\approx \phi _{m}[x-(n-m)d]$. This property is exactly verified
for a tilted lattice, and is a good approximation in the present
problem as long as the parabolic potential do not perturb too much the
translational invariance of the lattice wells over the spatial
extension corresponding to the atom wavepacket (that is, as long as
$V_0 \gg 1$ in our units).

At time $t=0$ the wavepacket $\Psi _{0}(x)$ is centered at $n_{0}$ and
has a spatial extension $\Delta n$: $\Psi _{0}=\sum_{n}c_{n}\phi_{n}(x)$.
At time $t$ one has: 
\begin{equation}
  \Psi (x,t)=\sum_{n}c_{n}\exp \left[ -i{\frac{n^{2}}{2}}t\right]
  \phi_{n}(x)
\label{eq:evolution}
\end{equation}
from which one can determine the dynamics of the system \cite{ref:WSbasis}.
For instance, the packet average position is
\begin{equation}
  \overline{x} = \langle \Psi | x | \Psi \rangle =
  \sum_{n}X_{m}c_{n}c_{n-m}^{\ast }\exp \left[ -im({n-} \frac{m}{2})t\right]
  \label{eq:xaverage}
\end{equation}
($X_{m}\equiv \left\langle \phi _{n}\left| x\right|
  \phi_{n+m}\right\rangle$ is almost independent of $n$).

Eq.~(\ref{eq:xaverage}) shows that, given the evolution of a
quantum system, its coherences $c_{n}c_{n-m}^{\ast }$ can be obtained from
the complex amplitude of the oscillations at the Bohr frequency
$\omega _{nm}$:
\begin{equation}
  \omega_{nm}=\epsilon _{n}-\epsilon _{m}=m(n-m/2) \text{ .}
  \label{eq:freqbohr}
  \end{equation}
The parabolic component of the potential produces
{\em non-degenerate} Bohr frequencies. In a tilted lattice
corresponding to a constant force $F$, the coherences between
neighbor sites $c_{n}c_{n-1}^{\ast }$ correspond to an oscillation
with the {\em degenerate} ($n$-independent) Bohr (or Bloch) frequency
$\omega _{B}=Fd/\hbar$ ($\omega_B=F$ in
reduced units). The parabolic potential, on the contrary, has
a local slope $F=x$ and the related ``local'' Bloch frequency
depends on $n$. The local Bloch frequency (LBF) is by definition
the Bohr frequency associated to neighbor sites at $n$ and $n-1$ ($m=1$): 
\begin{equation}
\omega _{B}(n)=\epsilon _{n}-\epsilon _{n-1}=n-\frac{1}{2} \text{.}
\label{eq:LocalFreq}
\end{equation}
The presence of such a frequency in the dynamics is a signature of the
coherence $c_{n}c_{n-1}^{\ast }$ that can be used
to track the wavepacket amplitude and phase, as shown below.

Experimentally, it is more convenient to work in the
momentum-representation, because the spatial amplitude of the BO is
in general very small whereas, in the momentum space, the amplitude of the
oscillation is of the order of
$\hbar/d$ (the extension of a Brillouin zone of the lattice),
which is easy to resolve with available experimental methods.
For cold atoms, for example, stimulated Raman
transitions \cite{ref:QuChaos} can be used to that end. 
A great deal of information can be obtained by just monitoring the
evolution of the probability for a given value $p$ of the momentum,
$P(p,t)$. We thus work in momentum space and
introduce the eigenstates $\varphi _{n}(p)$ in $p-$representation: 
\begin{eqnarray}
\varphi _{n}(p) &\approx &{\frac{1}{\sqrt{2\pi }}}\int \exp (ipx)\phi
_{n_{0}}\left[ x-(n-n_{0})\right] dx  \nonumber \\
&=&\exp \left[ i(n-n_{0})p\right] \varphi _{n_{0}}(p)
\label{eq:phi_p}
\end{eqnarray}
where we used the approximate translation invariance of the real-space
eigenfunctions $\phi_n(x)$. The evolution of the momentum
probability-distribution is then: 
\begin{eqnarray}
P(p,t) &=&|\Psi (p,t)|^{2}=|\varphi _{n_{0}}(p)|^{2}\times   \nonumber \\
&&\left( 1+\sum_{m\neq 0}\exp (imp)\right.   \nonumber \\
&&\left. \sum_{n\neq m}c_{n}c_{n-m}^{\ast }\exp \left[ -im\left( n-{\frac{m}{%
2}}\right) t\right] \right) \text{ .}
\label{eq:TheGoldenVero-QuentinEquation}
\end{eqnarray}

\begin{figure}
\includegraphics{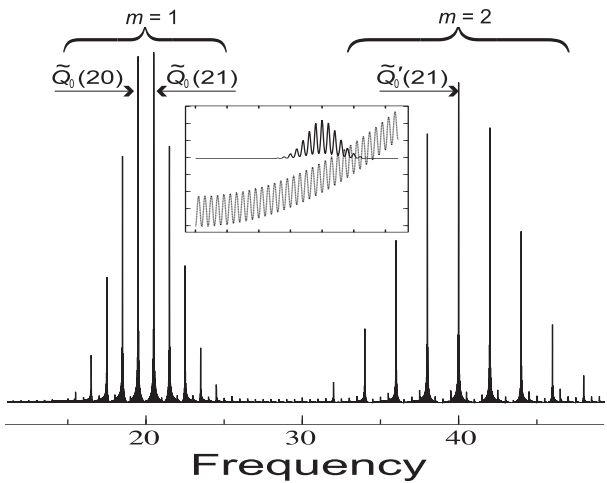}
\caption{\label{fig:BlochSpec}
Fourier transform of the probability of finding the system in the
zero-momentum state, $P_0(t)$.  The envelope of both frequency folds 
well reproduces the envelope of $|\Psi_0(x)|$ (see
inset). The arrows indicate the components corresponding to
the coherences between neighbor-sites
$\tilde{Q_0} (n)=c_nc^*_{n-1}$ ($n=20,21$) and next-to-neighbor-sites
$\tilde{Q_0}^\prime (21)=c_{21}c^*_{19}$.
$V_0= 90$, $n_0=21$, and $\Delta n=7$.}
\end{figure}

Fig.~\ref{fig:BlochSpec} displays the Fourier transform of the
quantity $P_0(t) \equiv P(0,t)$  and has been obtained as follows.
The time-dependent probability distribution $P_0(t)$ is
calculated by numerical integration of the  Schr{\"o}dinger equation
with the hamiltonian of Eq.~(\ref{eq:H}) and with the initial
wavepacket $\Psi_0(x)$ shown in 
the inset of Fig.~\ref{fig:BlochSpec} (the phase of $c_n$ is
arbitrarily set to $e^{in\pi/4}$). The spectrum of $P_0(t)$ presents
components at the Bohr frequencies contained in
Eq.~(\ref{eq:TheGoldenVero-QuentinEquation}), from which 
the amplitude and phase of the coherences $c_nc^*_{n-m}$ can
be deduced. The interpretation of the spectrum may be complicated
by the fact that Bohr
frequencies corresponding to energy differences higher than next-neighbor
wells may show up. These frequencies, corresponding to
$m>1$ in Eq.~(\ref{eq:freqbohr}), are higher than the
LBF [Eq.~(\ref{eq:LocalFreq})] typically by a factor $m$.
In the general case,
the spectrum is thus composed of ``folds'' described by an integer
value of $m$. However, a judicious choice of position of the
wavepacket in the parabolic potential allow to separate one
fold from the others. Frequency folds $m=1$\ and $m=2$\ are
resolved if $n_{0}+\Delta n/2+1/2<2\left( n_{0}-\Delta n/2-1\right) $
or, roughly, $\Delta n\ll n_{0}$. In producing Fig.~\ref{fig:BlochSpec},
we managed to satisfy this condition. The first frequency fold
is clearly seen as the
low-frequency part of the spectrum associated to the coherences $%
c_{n}c_{n-1}^{\ast }$. In Fig.~\ref{fig:BlochSpec}, the shape of the
wavepacket is reproduced by the envelopes of both the first
and the second frequency folds. This is the case if the wavepacket is
smooth enough at the scale of the lattice step that we can take
$|c_n| \approx |c_{n-1}|$, and then directly identify the amplitude of the
$n$-component of the spectrum with $|c_n|$.

The complete reconstruction of the spatial wavefunction
from the $c_{n}$ coefficients requires, in principle, the knowledge of the
eigenstates of system. The detailed spatial shape of the eigenstates
may depend on experimental parameters that may not
be known with a sufficient precision. With our system,
the shape of the localized states is almost independent of $n$, as it
is essentially determined by the lattice, the parabolic term giving
a small correction. As they are sharply localized, they can be
considered as delta-function-like, probing the local value of
the wavepacket. The detailed shape of the eigenstates is thus not
required for the reconstruction, at the a precision corresponding to
a lattice step.

We can now proceed to the complete reconstruction of a wavepacket.
We consider the first fold ($m=1$) of the spectrum, and the zero momentum
probability $P_{0}(t)$ only. The other folds have essentially the same
structure; other values of the probed momentum may be used to
extract complementary information or improve the
accuracy of the reconstruction. We thus reduce
Eq.~(\ref{eq:TheGoldenVero-QuentinEquation}) to the simpler form: 
\begin{eqnarray}
  Q_{0}(t) & \equiv &
  {\frac{P_{0}(t)}{|\varphi _{n_0}(0)|^{2}}}-1 \nonumber \\
  & = & \sum_{n}c_{n}c_{n-1}^{\ast }\exp
  \left[ -i\left( n-{\frac{1}{2}}\right) t \right] \text{ .}
\label{eq:SimplerFormOfGVQE}
\end{eqnarray}
One then sees that the Fourier component
$\tilde{Q_0}(n)=c_nc^*_{n-1}$ of $Q_{0}$ at frequency $n-1/2$ directly gives
the coherence between neighbor sites. If the wavepacket envelope is
smooth, the amplitudes can
be directly obtained by taking
$|c_{n}|^2\approx |\tilde{Q_0}(n)|$. If the phase also varies smoothly,
the phase shift from the site $n-1$ to the  site $n$ is also directly
obtained from the phase of $\tilde{Q_0}(n)$, up to an unavoidable,
but unimportant, global phase factor.

A more precise determination of the coefficients $c_{n}$ is obtained
by using the second frequency fold  
which, as one can easily convince oneself, gives the coherences
$c_n c^*_{n-2} \equiv  \tilde{Q_0^\prime}(n)$, and the identity:
\begin{equation}
\frac{\tilde{Q_0}(n)\tilde{Q_0}(n+1)}{\tilde{Q^\prime_0}(n+1)} =
|c_n|^2 \text{ .}
  \label{eq:reconstruction}
\end{equation}

\begin{figure}
\includegraphics{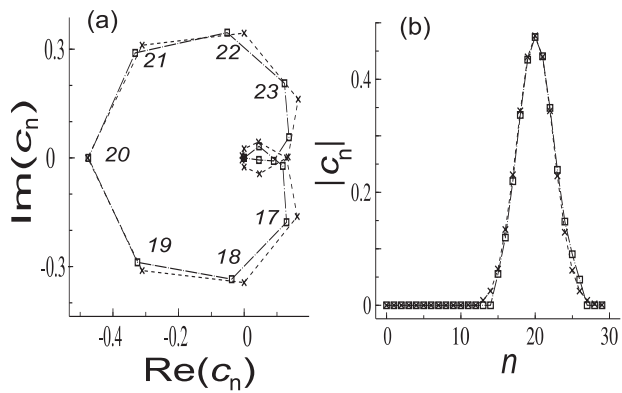}
  \caption{\label{fig:Wavepacket}
    Reconstruction of the initial wavepacket. Real and imaginary parts of
    $c_n$ are shown in (a), evidencing the very good agreement between
    phases and amplitudes of the original wavepacket components
    ($\times$) and the reconstructed values ($\square$). The method
    correctly reproduces the phase mismatch between neighbors
    $\phi=\pi/4$. Curve (b) shows explicitly the amplitudes
    versus $n$ and compares these values with the ones of the original
    wavepacket.}
\end{figure}

The phase shifts can be straightforwardly obtained from the 
$\tilde{Q_0}(n)$.  We display
in Fig.~\ref{fig:Wavepacket} the coherences $c_n$
deduced from the spectrum of Fig.~\ref{fig:BlochSpec}. The agreement 
with the amplitudes of the original wavepacket is excellent. The
figure also shows real and imaginary parts of the complex
amplitudes in very good agreement with the original wavepacket.

We numerically verified that the above method can be applied
to wavepackets of arbitrary shape. The reconstructed state then
reproduces the wavepacket at the scale of the lattice step, all finer features
are lost. A large number of delocalized states can appear in the
decomposition of the initial state, but their contribution to the
Bloch-oscillation spectrum is to add a wide background whose 
amplitude is much smaller than the peaks associated to the
coherence between localized states.

Consider now the case in which there are many atoms simultaneously
interacting with the potential. If the  density of atoms is low, so 
that there is in average one atom per site (or less), the present method
reconstructs the atom distribution. If the atomic density is
higher, the method measures the ensemble average of the
atomic correlation $\langle c_{n}c_{n-m}^{\ast }\rangle $.
This quantity is very important because
it determines the ability of the system to display coherent quantum
dynamics \cite{ref:WSbasis}. Finally, note that the present
method can also be applied to a tridimensional lattice: adding a
parabolic potential in a given direction of the lattice
will probe the wavefunction along this direction.

In conclusion, we have presented a method allowing reconstruction of the
spatial wavefunction of a quantum system using the Fourier spectrum of the
Bloch oscillations  in a periodic lattice superposed
to a parabolic potential. Our method is quite general: it does not depend,
e.g. on internal properties of the quantum particle being studied, or on the
shape of the lattice wells. A parabolic shape is not
necessary provided that (i) the Bohr (or Bloch) frequencies of the
system are not degenerated and (ii) the approximate translational symmetry of
the localized states is preserved.
The method requires a potential shape presenting both a
non-constant and non-vanishing slope in the region of interest:
the local slope induces local Bloch
oscillation and the spatial dependence, in turn, leads to  the
position-sensitivity of the method.
In principle, the method is applicable to a variety of
systems, including cold atoms in light potentials, Bose-Einstein
condensates and electrons in an
adequately constructed superlattice. Applied to a Bose-Einstein
condensate, whose Bloch oscillations have been recently
observed \cite{ref:BECBO}, the present method may be used to probe
nonlinear collective interactions. The method is also able to 
probe the {\em spatial
entanglement} of two or more atoms, which can be of interest in the
context of quantum information and quantum computing. We shall
treat these interesting applications in the future.

The authors would like to acknowledge D. Delande,
P. Szriftgiser and H. Lignier for fruitful discussions. This work is
partially supported by an action ``ACI Photonique'' of the Minist\`{e}re de
la Recherche. Laboratoire de Physique des Lasers, Atomes et Mol{\'e}cules 
(PhLAM) is UMR 8523 du CNRS et de l'Universit{\'e} des Sciences et
Technologies de Lille. Centre d'Etudes et Recherches Lasers et Applications
(CERLA) is FR 2416 du CNRS, supported by Minist\`{e}re de la
Recherche, R\'{e}gion Nord-Pas de Calais and Fonds Europ\'{e}en de
D\'{e}veloppement Economique des R\'{e}gions (FEDER).


\begin{thebibliography}{99}
\bibitem{note:quasi-degeneracy} The energy difference
  $|\delta _{n}^{A}-\delta _{n}^{S}|/n^2$ is typically of the
order of $10^{-5}$ for $V_0=90$. Moreover, the $\delta_n^{A,S}$ 
are roughly independent of $n$.
\end{thebibliography}

\end{document}